\documentclass{cimento}
\usepackage{graphicx}
\title{Insight into Emergence of Hadron Mass from $\boldmath N^*$ Electroexcitation Amplitudes}
\author{V.I.~Mokeev\from{ins:jlab},
D.S.~Carman\from{ins:jlab}
}
\instlist{\inst{ins:jlab} Thomas Jefferson National Accelerator Facility, Newport News, Virginia 23606, USA
}

\begin{document}

\maketitle

\begin{abstract}

The emergence of hadron mass represents one of the most challenging and still open problems in contemporary hadron physics. The results 
on the nucleon resonance electroexcitation amplitudes available from the CLAS data on $\pi N$ and $\pi^+\pi^-p$ electroproduction analyzed
within the continuum Schwinger method open up a new avenue for gaining insight into the strong interaction dynamics that are responsible 
for the generation of the dominant part of hadron mass. Future prospects of these studies in experiments of the 12-GeV era with CLAS12 
and after a potential increase of the CEBAF energy up to 22 GeV will offer a unique opportunity to explore the full range of distances 
where the dominant part of hadron mass and $N^*$ structure emerge from QCD.  

\end{abstract}

\section{Introduction}

An extensive research program is now in progress in Hall~B at Jefferson Lab with the objective to determine the nucleon resonance ($N^*$)
electroexcitation amplitudes, also known as the $\gamma_vpN^*$ electrocouplings, for most states in the mass range $W$ up to 2.0~GeV, 
for photon virtualities $Q^2 < 5$~GeV$^2$ from the data on the exclusive meson electroproduction channels measured with CLAS and 
analyzed within different reaction models~\cite{Mokeev:2022xfo,Carman:2020qmb,Aznauryan:2011qj}. This information will be further extended
towards $Q^2$ up to 10~GeV$^2$ from the data expected in the measurements with the CLAS12 detector that are currently in progress
\cite{Proceedings:2020fyd,Burkert:2018nvj}. These studies offer unique insight into many facets of the strong interaction dynamics in the
regime of large (comparable with unity) QCD running coupling $\alpha_s/\pi$ seen in the generation of $N^*$ states of different quantum
numbers with different structural features~\cite{Burkert:2017djo,Burkert:2019kxy,Giannini:2015zia,Aznauryan:2017nkz,Aznauryan:2012ba,
Obukhovsky:2011sc,Obukhovsky:2019xrs,Lyubovitskij:2020gjz}. Analyses of the results on the $Q^2$-evolution of the $\gamma_vpN^*$
electrocouplings within continuum Schwinger methods (CSM) that provide a connection to the QCD Lagrangian, open up a promising new avenue 
in the exploration of the strong interaction dynamics responsible for the emergence of the dominant part of hadron mass
\cite{Carman:2023zke,Ding:2022ows}.  

\begin{figure}[htbp] 
\begin{center}
\includegraphics[width=5.0cm]{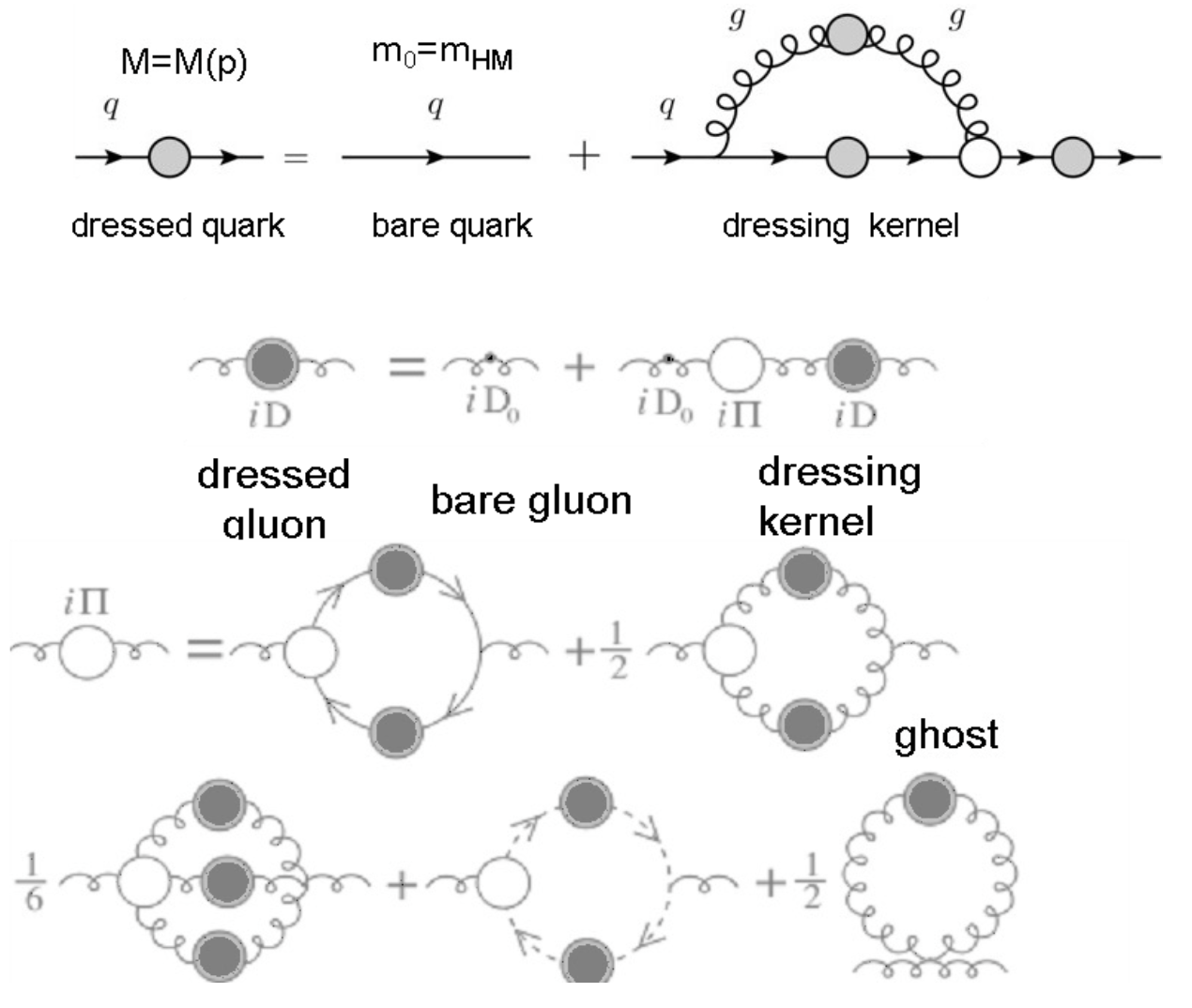}
\hspace{3.0mm}
\includegraphics[width=7.0cm]{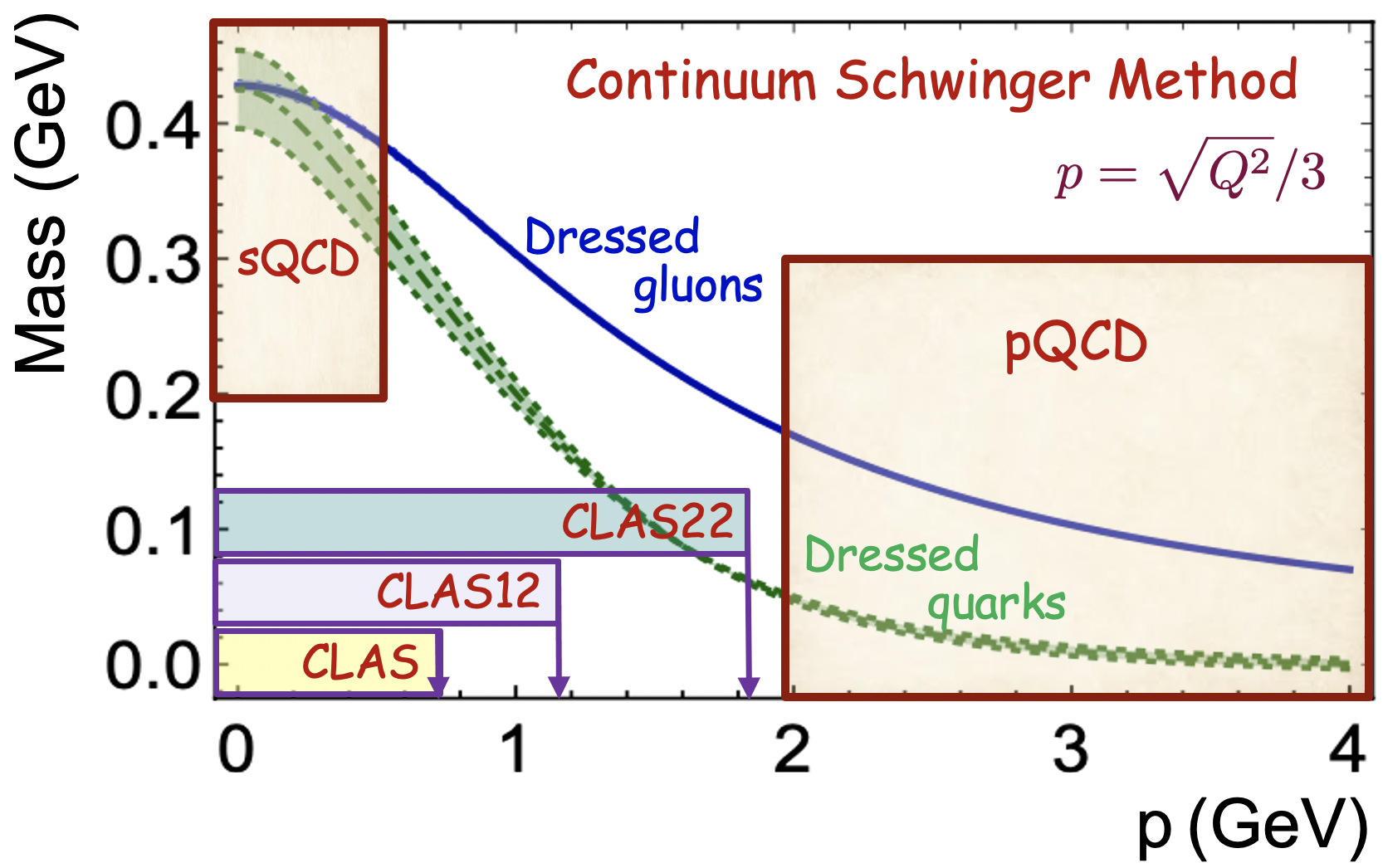}
\caption{Concepts underlying emergent hadron mass (EHM) within the continuum Schwinger method (CSM)~\cite{Ding:2022ows,Roberts:2021nhw}: 
QCD-driven bare quark and gluon dressing (left) and the momentum dependencies of the dressed quark and gluon masses
(right)~\cite{Roberts:2021xnz}.} 
\label{EHM_CSM}
\end{center}
\end{figure}

\section{CSM Concept for Emergent Hadron Mass}

The emergence of hadron mass is the most challenging open problem in hadron physics. This problem is revealed through the comparison 
between the observed nucleon masses and the sum of the masses of their quark constituents. The sum of the three bare $u$ and $d$  
quark masses~\cite{ParticleDataGroup:2022pth} that enter into the QCD Lagrangian and generated through the Higgs mechanism accounts for 
$<$2\% of the measured nucleon masses. This observation suggests that the dominant part of hadron mass is generated in processes other 
than the Higgs mechanism.

The QCD gauge gluons of zero mass should be dressed in the processes imposed by the interaction part of the QCD Lagrangian shown in the 
bottom left panel of Fig.~\Ref{EHM_CSM}. The gluon self-interaction creates dynamically generated gluon mass that emerges in the 
transverse part of the gluon vacuum polarization tensor $P_{\mu \nu}$ that preserves gauge invariance~\cite{Ding:2022ows,Roberts:2021nhw}.
The scalar function for the gluon vacuum polarization tensor $P(p^2)$ behaves as $\frac{m_g^2(p^2)}{p^2}$, where $p^2$ is the squared 
four-momentum of the dressed gluon. Consequently, the dressed gluon propagator $\propto \frac{1}{p^2(1+P(p^2))} = \frac{1}{p^2+m_g^2(p^2)}$, demonstrating the emergence of non-zero running mass $m_g(p^2)$ of the dressed gluon. The gluon self-interaction
also causes a rapid increase of the QCD running coupling $\alpha_s/\pi$ in the transition from the perturbative to the strongly coupled
regimes. Furthermore, at gluon momenta $p < 0.5$~GeV, the coupling $\alpha_s/\pi$ becomes almost momentum independent owing 
to the finite value of the gluon mass as $p\to 0$~\cite{Cui:2019dwv}. The bare QCD quark should be dressed in the processes shown in the
top left panel in Fig.~\ref{EHM_CSM}. Its propagator behaves $\propto \frac{1}{A(p^2)p^2+B(p^2)} = \frac{Z}{p^2+(B(p^2)/A(p^2))}$, 
demonstrating the dynamical generation of the dressed quark mass $m_q^2=B(p^2)/A(p^2)$ running with $p^2$, the dressed quark squared 
four-momentum, and the renormalization parameter $Z$ for the quark wave function. When the QCD running coupling $\alpha_s/\pi$ becomes 
comparable with unity ($\alpha_s/\pi > 0.3$), the energy stored in the gluon field is transformed into a momentum dependence of the 
dressed quark masses as shown in the right panel of Fig.~\ref{EHM_CSM}. The momentum dependencies of the dressed quark and gluon masses 
have been computed within CSM as the solution of the QCD equations of motion for the quark and gluon fields. The CSM predictions on the 
dressed quark/gluon mass functions have been confirmed in lQCD studies~\cite{Oliveira:2018lln}. The rapid increase seen in the momentum
dependence of the dressed quark mass when the quark momentum decreases underlies the emergence of the dominant part of the masses of the 
ground and excited states of the nucleon as bound systems of three dressed quarks. 

\section{Insight into EHM from Results on $Q^2$-evolution of $\gamma_vpN^*$ Electrocouplings}

The ground and excited states of the proton emerge as bound systems of three dressed quarks. In the processes of $N^*$ electroexcitation,
the virtual photon interacts with the dressed quarks. Since the running mass of the dressed quark is encoded in dressed quark
propagator, the information on the $Q^2$ evolution of the $\gamma_vpN^*$ electrocouplings allows us to map out the momentum dependence 
of the dressed quark mass.

Analyses of the CLAS data on $\pi N$, $\eta N$, and $\pi^+\pi^-p$ electroproduction have provided the only available results on the 
$Q^2$ evolution of the $\gamma_vpN^*$ electrocouplings for most $N^*$ states in the mass range up to 1.8~GeV for $Q^2 < 5$~GeV$^2$
\cite{Mokeev:2022xfo,Carman:2020qmb,Burkert:2022ioj,HillerBlin:2019jgp}. Consistent results on the $\gamma_vpN^*$ electrocouplings of 
the resonances in the mass range up to 1.6~GeV available from independent studies of $\pi N$ and $\pi^+\pi^-p$ electroproduction with
different non-resonant contributions demonstrated the capabilities of the reaction models~\cite{Aznauryan:2009mx,Mokeev:2012vsa} 
developed by the CLAS Collaboration for the credible extraction of these quantities.

\begin{figure}[htbp] 
\begin{center}
\includegraphics[width=13.5cm]{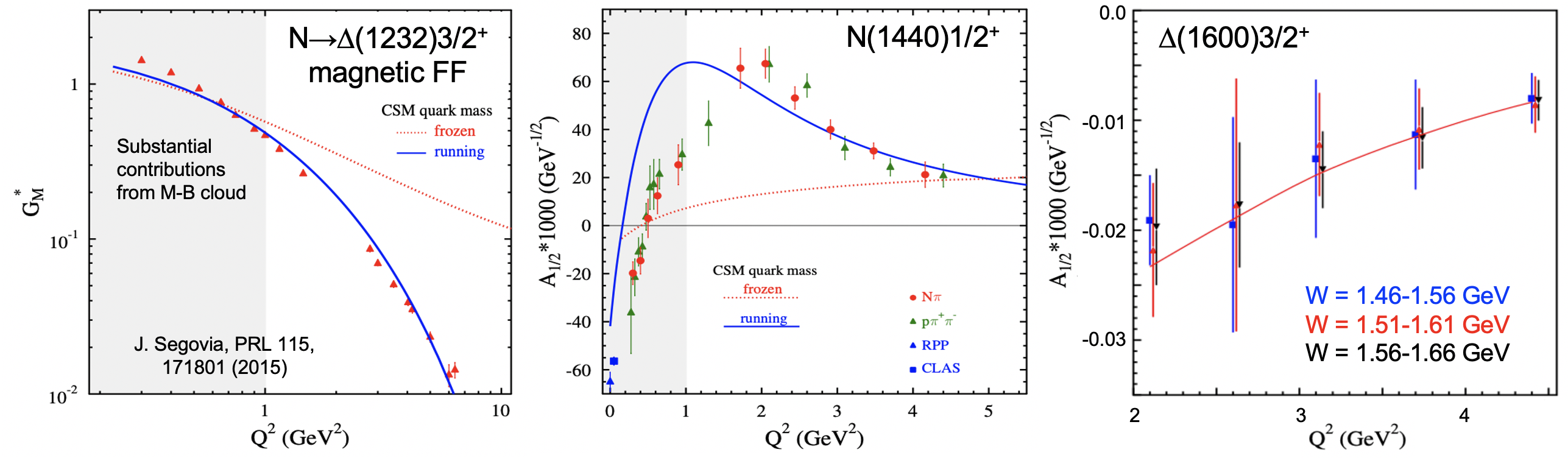}
\caption{Description of the CLAS results on the $N \to \Delta$ magnetic transition form factor (left)~\cite{Aznauryan:2009mx} and
$N(1440)1/2^+$ electrocouplings~\cite{Aznauryan:2009mx,Mokeev:2012vsa,Mokeev:2023zhq} (middle) achieved within CSM
\cite{Segovia:2014aza,Segovia:2015hra} by employing the same dressed quark mass function deduced from the QCD Lagrangian (blue solid 
lines). The CSM evaluations with momentum-independent (frozen) quark mass are shown by the red dashed lines~\cite{Segovia:2014aza,
Wilson:2011aa}. (Right) CLAS results on the $\Delta(1600)3/2^+$ electrocouplings obtained from analyses of $\pi^+\pi^-p$ electroproduction
cross sections within $W$-intervals as indicated for $Q^2$ from 2.0 to 5.0~GeV$^2$~\cite{Mokeev:2023zhq}. The CSM predictions are shown 
by the red line~\cite{Lu:2019bjs}.} 
\label{RES_CSM}
\end{center}
\end{figure}

The experimental results on the $Q^2$-evolution of the dominant $N \to \Delta$ magnetic transition form factor and all electrocouplings 
of the $N(1440)1/2^+$ have been described successfully within CSM~\cite{Segovia:2014aza,Segovia:2015hra} starting from the QCD Lagrangian
for $Q^2 > 0.8$~GeV$^2$ for the $\Delta(1232)3/2^+$ and for $Q^2 > 2.0$~GeV$^2$ for the $N(1440)1/2^+$. Within these $Q^2$ ranges the core
of three dressed quarks becomes the biggest contributor to the structure of these states, allowing for direct comparison between the data
and the CSM predictions that account for the contribution from only the quark core. A good description of the electrocouplings of these
resonances of different structure (spin-isospin flip and the first radial excitation of three dressed quarks for the $\Delta(1232)3/2^+$ 
and $N(1440)1/2^+$, respectively) shown in the left and the middle panels in Fig~\ref{RES_CSM}, was achieved with the same dressed quark 
mass function deduced from the QCD Lagrangian (see Fig.~\ref{EHM_CSM} right), that was already used in the successful description of the
pion and nucleon elastic electromagnetic form factors. This success offers strong evidence for dressed quarks with running mass as the 
active degrees of freedom in the structure of the pion, ground states of the nucleon, and the $\Delta(1232)3/2^+$ and $N(1440)1/2^+$, 
and demonstrates the capability for mapping out the momentum dependence of the dressed quark mass from the results on the $Q^2$-evolution 
of the $\gamma_vpN^*$ electrocouplings.

The electrocouplings of the $\Delta(1600)3/2^+$ were predicted for the $Q^2$-range from 2.0 to 5.0~GeV$^2$ within the CSM approach in
2019 with the dressed quark mass function described above~\cite{Lu:2019bjs}, when experimental results were not available. The experimental
results on the $\Delta(1600)3/2^+$ electrocouplings were obtained in 2023~\cite{Mokeev:2023zhq} from analyses of the nine independent 
one-fold differential $\pi^+\pi^-p$ electroproduction cross sections measured with CLAS~\cite{CLAS:2017fja,Trivedi:2018rgo} in three 
overlapping $W$-intervals for $Q^2$ from 2.0 to 5 0~GeV$^2$ (see Fig.~\ref{RES_CSM} right). The $\Delta(1600)3/2^+$ contributes 
substantially into all three $W$ intervals with different non-resonant backgrounds. The $\Delta(1600)3/2^+$ electrocouplings extracted 
from the data in the three $W$-intervals are consistent, demonstrating credible extraction of these quantities. These experimental 
results have confirmed the CSM predictions, which solidifies evidence for insight into the dressed quark mass function.

The $Q^2$-range for the $\gamma_vpN^*$ electrocouplings covered by the meson electroproduction data from CLAS spans up to 5~GeV$^2$,
allowing us to map out the range of quark momenta where $<$30\% of hadron mass is expected to be generated (see Fig.~\ref{EHM_CSM} right).
CLAS12 is the only available facility capable of extending the $Q^2$-coverage of the results on the $\gamma_vpN^*$ electrocouplings up to
10~GeV$^2$ from measurements of $\pi N$, $KY$, and $\pi^+\pi^-p$ electroproduction~\cite{Mokeev:2022xfo}. The foreseen CLAS12 results will
allow us to explore the range of distances where $\approx$50\% of hadron mass is generated. A further increase of the CEBAF energy up to 
22~GeV~\cite{Accardi:2023chb} will make it possible to explore the $\gamma_vpN^*$ electrocouplings over a still broader $Q^2$-range.
Simulation studies performed at 22~GeV with CLAS12 suggest that luminosities up to $2-5 \times 10^{35}$~cm$^{-2}$s$^{-1}$ would be
sufficient in order to obtain the $\gamma_vpN^*$ electrocouplings for $Q^2$ up to 30~GeV$^2$. These results will offer the only foreseen
opportunity to explore the full range of distances where the dominant part of hadron mass and $N^*$ structure emerge from QCD (see 
Fig.~\ref{EHM_CSM}), addressing the key open problem of the Standard Model on the emergence of hadron mass.

\section{Conclusions and Outlook}

Nucleons and their excited states are the most fundamental three-body systems in Nature. If we do not understand how QCD builds each state
in the complete $N^*$ spectrum, then our understanding of the strongly coupled QCD regime remains incomplete. High-quality meson
electroproduction data from the 6-GeV era at JLab have allowed for the determination of the electrocouplings of most $N^*$s in the mass
range up to 1.8~GeV for $Q^2 < 7.5$~GeV$^2$. A good description of the $\Delta(1232)3/2^+$, $N(1440)1/2^+$, and $\Delta(1600)3/2^+$
electrocouplings for $Q^2 < 5$~GeV$^2$ achieved within CSM with the same dressed quark mass function inferred from the QCD Lagrangian 
and used in the successful description of the elastic pion and nucleon electromagnetic form factors, offers sound evidence for insight 
into the momentum dependence of the dressed quark mass. Extension of the results on the $\gamma_vpN^*$ electrocouplings into the $Q^2$ 
range from 5 - 30~GeV$^2$ from the CLAS12 measurements in the 12-GeV era and after the increase of the CEBAF energy up to 22 GeV,
will offer the only foreseen opportunity to explore how the dominant part of hadron mass and $N^*$ structure emerge from QCD. 

\section{Acknowledgments}
This work was supported in part by the U.S. Department of Energy (DOE) under Contract No. DE-AC05-06OR23177 and Jefferson Science 
Associates (JSA).

\bibliography{References}



\end{document}